\documentclass[]{article}
\usepackage{spconf,amsmath,graphicx,amssymb,delarray,subfigure,verbatim,booktabs}
\usepackage{diagbox, makecell}
\usepackage{multirow}
\usepackage[english]{babel}
\usepackage{lipsum}
\usepackage{bbding}
\usepackage{setspace}

\title{BAE-Net: A Low complexity and high fidelity Bandwidth-Adaptive neural network for speech super-resolution}
\name{\makecell[c]{Guochen Yu$^{\star \ast}$, Xiguang Zheng$^{\star \ast}$\thanks{Contributed equally to this work as co-first authors. Chengshi Zheng is the corresponding author.}, Nan Li$^{\star}$, Runqiang Han$^{\star}$, Chengshi Zheng$^{\dagger}$\\
Chen Zhang$^{\star}$, Chao Zhou$^{\star}$, Qi Huang$^{\star}$, Bing Yu$^{\star}$}} 

\address{$^{\star}$ Kuaishou Technology, Beijing, China\\
    $^{\dagger}$ Institute of Acoustics, Chinese Academy of Sciences, Beijing, China\\
}

\begin{document}
\ninept
\maketitle
\begin{abstract}
 Speech bandwidth extension (BWE) has demonstrated promising performance in enhancing the perceptual speech quality in real communication systems. Most existing BWE researches primarily focus on fixed upsampling ratios, disregarding the fact that the effective bandwidth of captured audio may fluctuate frequently due to various capturing devices and transmission conditions. In this paper, we propose a novel streaming adaptive bandwidth extension solution dubbed BAE-Net, which is suitable to handle the low-resolution speech with unknown and varying effective bandwidth. To address the challenges of recovering both the high-frequency magnitude and phase speech content blindly, we devise a dual-stream architecture that incorporates the magnitude inpainting and phase refinement. For potential applications on edge devices, this paper also introduces BAE-NET-lite, which is a lightweight, streaming and efficient framework. Quantitative results demonstrate the superiority of BAE-Net in terms of both performance and computational efficiency when compared with existing state-of-the-art BWE methods.
    \end{abstract}
    \vspace{-1mm}
    \section{Introduction}

 In real-time communication (RTC) scenarios, bandwidth extension (BWE), also known as audio super-resolution, is often employed to recover the high-resolution (HR) speech from its corresponding low-resolution (LR) input, where the missing HR part is commonly caused by acquisition devices and transmission. The main purpose of BWE is to increase the naturalness and clarity of speech with arbitrary sampling rate~{\cite{antons2012too}}.
 
 
 Conventional signal processing based BWE methods typically include linear predictive coding (LPC) analysis~{\cite{soong1993optimal}}, Gaussian mixture models (GMMs)~{\cite{park2000narrowband}}, and Hidden Markov Model (HMM)~{\cite{jax2003artificial}}. Over the past few years, a plethora of deep learning based BWE researches have triggered breakthroughs, which can be roughly categorized into spectrum-based methods{\cite{li2015deep, li2018speech, schmidt2018blind, eskimez2019adversarial,hu2020phase,mandel2023aero}} and waveform-based methods~{\cite{kuleshov2017audio, kim2019bandwidth, kumar2020nu, hao2020time,li2021real,su2021bandwidth, lee2021nu, han2022nu}}. Typical spectrum-based approaches reconstruct the high-frequency components from the corresponding low-frequency spectral representations such as magnitude spectrum and the log-power magnitude spectrum. Generally, the low-frequency phase information is  directly flipped as the high-frequency phase to reconstruct the time-domain HR waveform, which hinders further performance improvements{\cite{li2015deep,li2018speech,eskimez2019adversarial}}. Recently, a novel spectrum-domain based BWE method dubbed AERO directly operates on the complex-valued spectrum~{\cite{mandel2023aero}}, aiming to recover high-frequency magnitude and phase information simultaneously, and thus achieves better performance when compared with the magnitude-only methods. The waveform-based approaches typically employ the spline interpolation and encoder-decoder networks to learn the LR to HR mapping in the time domain~{\cite{kumar2020nu, hao2020time,li2021real,su2021bandwidth, lee2021nu, han2022nu}}.
The main challenge for time-domain methods is that modeling the high-resolution waveform directly causes expensive computational demands. 

\begin{figure}[t]
    \centering
    \centerline{\includegraphics[width=0.9\columnwidth]{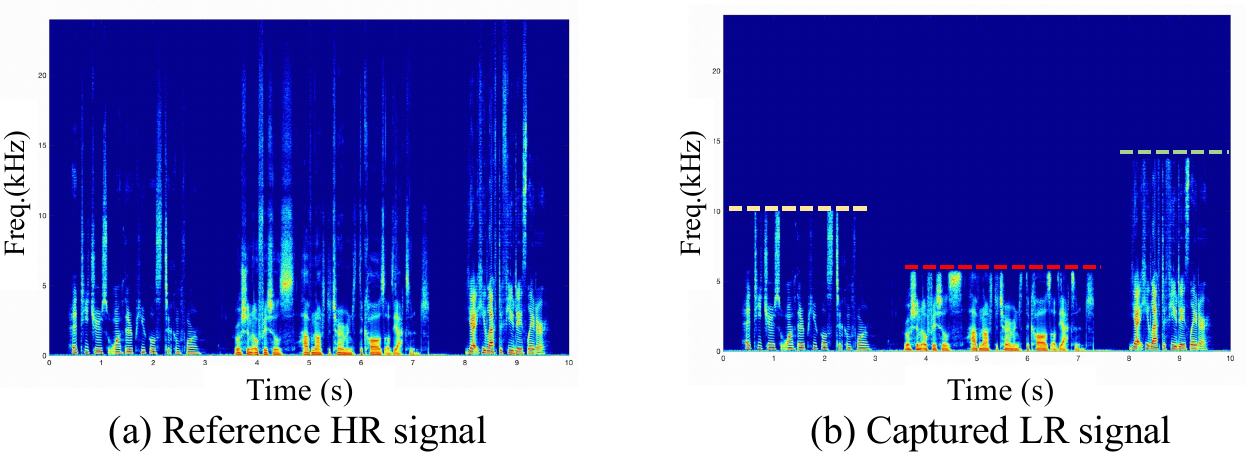}}
    \vspace{-0.3cm}
    \caption{Visualization of spectrograms of full-band reference HR speech and bandwidth-fluctuated LR speech}
    \label{fig:example}
    \vspace{-0.6cm}
    \end{figure}

Admittedly, existing deep learning based methods have achieved significant improvements on perceptual speech quality, at least two intrinsic problems remain challenging for practical RTC applications.
The first challenge is that the effective bandwidth of the captured audio frequently fluctuates caused by variable reasons in RTC systems. 
For example, different mobile devices may have different intrinsic capturing sample rates. In noisy environments, the speech enhancement algorithms may undesirably erase the high-frequency speech components during the low SNR periods, while preserving the high-frequency speech components when the SNR becomes high. Different transmission conditions may also affect the effective bandwidth when decreasing the encoding bitrate under severe upstream packet loss and jitter conditions. Figure~{\ref{fig:example}} illustrates an example of realistic RTC capture of the fluctuated effective bandwidth. Unfortunately, most existing approaches only support predefined and fixed up-sampling scales such as 8 to 16kHz and 8 to 24kHz sampling rate.
In addition, although several BWE researches such as NU-wave2~{\cite{han2022nu}} support the arbitrary sampling rate, the complexity and latency of the existing fixed and arbitrary sampling rate based BWE methods is extremely difficult to be deployed on the mobile devices for real-time applications~{\cite{li2021real}}.



\begin{figure*}[ht!]
    \centering
    \centerline{\includegraphics[width=1.6\columnwidth]{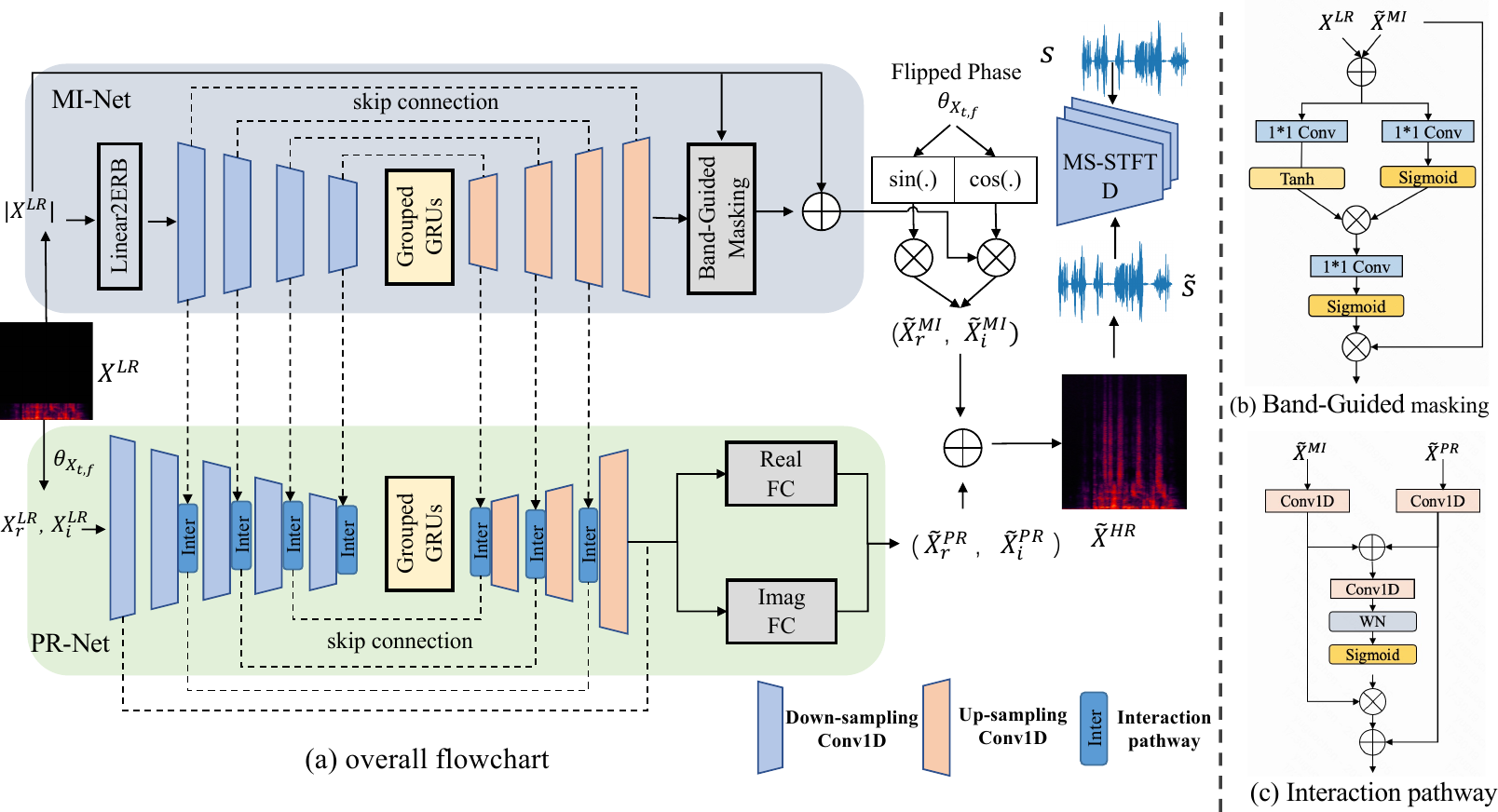}}
    \vspace{-0.4cm}
    \caption{(a) The flowchart of the proposed BAE-Net. (b) Band-Guided Masking module. (c) The information interaction pathway.}
    \label{fig:diagram-system}
\vspace{-0.6cm}   
\end{figure*}

 In this paper, we propose a low-complexity and streaming \textbf{B}andwidth-\textbf{A}daptive \textbf{E}xtension neural network (\textbf{BAE-Net}) in the spectral domain to alleviate the above-mentioned limitations. 
The major contributions in this work can be summarized as follows. \textbf{(i)} The proposed dual-stream network consisting of a magnitude inpainting network (MI-Net) and a phase refinement network (PR-Net), to facilitate both magnitude and phase information recovery. Specifically, a lightweight MI-Net is designed to inpaint the high-frequency magnitude components from speech signals with variable effective bandwidth and PR-Net aims to implicitly refine the flipped phase from a complementary perspective. Additionally, a novel information interaction pathway is introduced to leverage the information from MI-Net which can effectively resolve the difficulty of independent phase prediction. \textbf{(ii)} For edge devices, we introduce a lite version of BAE-Net called \textbf{BAE-Net-lite}, which focuses on magnitude inpainting only and employs the flipped phase to reconstruct waveform. BAE-Net-lite achieves comparable performance to baselines while significantly reducing network parameters (\textbf{500 K}) and computational complexity (\textbf{0.05 GMACs}). Therefore, BAE-Net-lite is potentially applied in real-time communication applications, such as live video/audio streaming and remote conferencing systems.

To validate our claims, we conduct extensive experiments on both fixed upsampling ratios (i.e., 8-48kHz, 16-48kHz sampling rate) and fluctuated bandwidth speech. Objective tests demonstrate that BAE-Net achieves comparable performance to several state-of-the-art baselines on fixed upsampling scales and exhibits robustness in real bandwidth-unknown scenarios, while BAE-Net-lite incurs dramatically less computation cost.

\vspace{-0.4cm}    
    \section{Methodology\label{Section2}}
 \vspace{-0.4cm}
    \label{Sec2}
        \subsection{ Overview\label{Section21}}
        \vspace{-0.1cm}
 
    The overall diagram of the proposed system is illustrated in Figure~{\ref{fig:diagram-system}} (a), which is mainly comprised of two parallel streams, namely a magnitude inpainting network (MI-Net) and a phase refinement network (PR-Net).
    By virtue of recent studies in decoupling-style phase-aware speech enhancement~{\cite{yin2020phasen,li2021simultaneous,yu2021dual, zheng2023sixty}}, the basic idea of BAE-Net is to decouple the prediction of HR magnitude spectrum and phase information. As shown in Figure~{\ref{fig:diagram-system}}, the input feature of BAE-Net is the real and imaginary parts of the 48kHz-sampling-rate speech STFT spectrum with fluctuated effective bandwidth, denote as $X^{LR}\in \mathbb{R}^{T\times F\times 2}$, where $T$ represents the number of frames and $F$ represents the number of frequency bands. Then, we decouple the STFT complex-valued spectrum into magnitude and phase, and feed the LR magnitude spectrum $\vert{X}^{LR}\rvert$ into MI-Net. 
    
    
Instead of only generating the high-frequency spectral features which requires an accurate effective bandwidth detection scheme, MI-Net is designed to estimate full-band magnitude spectrum, which contains the existing low-frequency and the high-frequency components. After estimating the inpainted magnitude spectrum, we utilize the flipped phase to recover the coarsely estimated HR complex spectrum. Specifically, the flipped phase of the high-frequency range is manually produced by flipping the phase of the narrow-band speech with 8 kHz sampling rate (4 kHZ effective bandwidth) and adding a negative sign until it covers the full frequency range (48 kHz sampling rate).
To refine the flipped phase, we employ PR-Net to estimate the full-band residual real and imaginary (RI) components. The input feature of PR-Net is the full-band complex spectrum, where the RI components are stacked along the frequency axis. Note that the phase itself is highly unstructured and hard to estimate, we propose an information interaction pathway to leverage the information from MI-Net to guide the phase prediction. After up-sampled Conv1Ds, two fully connected (FC) layers are employed to predict the residual HR RI components (i.e., $(\widetilde{X}^{PR}_{r}, \widetilde{X}^{PR}_{i})$) separately, thus refining the phase information. Finally, the predicted HR spectrum can be formulated as the summation of the predicted component by MI-Net and that by PR-Net.

 \vspace{-0.3cm}   
    \subsection{ Low-complexity Magnitude Inpainting Network\label{Section23}}
 \vspace{-0.2cm}   
    In the magnitude spectrum inpainting stream, we design a light-weight network MI-Net to derive the full-band HR magnitude spectrum. Unlike previous studies that only fill the high-frequency missing components~{\cite{li2015deep,li2018speech,eskimez2019adversarial}}, or recent studies that first perform upsampling and then estimate the whole low-frequency and high-frequency components~{\cite{hao2020time,kumar2020nu, han2022nu}}, we aim to slightly refine the existing low-frequency components and fill the missing high-frequency components. As illustrated in Figure~{\ref{fig:diagram-system}}, to reduce the redundancy of features at ineffective high-frequencies and decrease the computational burden, we first rearrange the 769-D linear STFT spectrum into 128 spectral bands with a triangular equivalent rectangular bandwidth (ERB) filter bank. Four down-sampled 1D convolutional layers (Conv1D) with weight normalization~{\cite{salimans2016weight}} is employed to compressed the ERB-scaled spectrum, while the number of channel is set to $\left\{128,128,64,64\right\}$ to reduce the frequency size and the kernel size is set to 3 in all convolutions. For streaming inference, two frame buffer is adopted to ensure convolution continuity across each frames. Four symmetrical up-sampled Conv1D are adopted to reconstruct the HR spectrum with summation skip connections, where the number of channel is set to $\left\{64,128,128,769\right\}$ and the kernel size is set to 3. Within down-sampling and up-sampling layers, two lightweight stacked grouped GRU are inserted into down-sampled and up-sampled layers with the group number set to 4. 

    To avoid modifying too much on low-frequency components by the mapping function, we propose a band-guided masking module, as illustrated in Figure~{\ref{fig:diagram-system}}(b). The band-guided masking module receives both original LR input $\vert{X}^{LR}\rvert$ and the output of the last up-sampled layer (i.e., $\vert{X}^{UP}\rvert$), and derives a gain function $\mathcal{G}^{MI}$ by a dual-path gating mechanism, which is then element-wise multiplied with the pre-estimated full-band spectrum to automatically filter and preserve different frequency range. 
    The final output of MI-Net is the summation of the input and estimated HR magnitude spectra.
    
        
    \vspace{-0.3cm} 
    
    \subsection{Phase Refinement Network \label{Section21}}
    \vspace{-0.2cm}
        In the phase-refinement stream, PR-Net aims to correct the phase distribution as well as further recovering the magnitude spectrum from a complementary aspect. Given the original RI components $\left\{ X_{r}, X_{i}\right\}$ of the low-resolution speech as the input, 
        PR-Net is designed to estimate the residual complex spectral details. Similar to MI-Net, five grouped down-sampling convolutional layers are first employed to reduce the frequency size, with the group number setting to $\left\{2,2,2,1,1\right\}$. The number of channel is set to $\left\{512, 128,128,64,64\right\}$ in the frequency axis, and the kernel size is set to 3 in all convolutions.
        
        Considering the difficulty in modeling phase modeling individually, an information interaction layer is introduced to leverage the information from the magnitude stream to guide the phase rectification. Figure{\ref{fig:diagram-system}}(c) shows the detailed flowchart of the proposed interaction pathway. Taking the interaction pathway in down-sampling layers as an example, we first merge the intermediate feature from MI-Net with that of the previous down-sampling layer in PR-Net. Then, the summation is fed into a mask module to derive a mask, which aims to automatically filter out the magnitude-related feature from MI-Net. Finally, the filtered feature is added with the intermediate phase-related feature in PR-Net to get the interacted feature. 

        Following the down-sampling layers and grouped GRUs, three up-sampling 1D convolutions are adopted to enlarge the frequency size and the number of channel is set to $\left\{128, 128, 512\right\}$. After up-sampling layers, two FC layers are employed to reconstruct the residual RI components (i.e., $({X}^{PR}_{r}, {X}^{PR}_{i})$) in parallel, and they are then merged with the coarsely inpainted complex spectrum by MI-Net to derive the final HR spectrum.

\vspace{-0.3cm}
    \subsection{Training Objective\label{Section24}}
    \vspace{-2mm} 
     The training loss is a weighted sum of the reconstruction loss term and adversarial loss term. For the reconstruction loss, we first employ a waveform loss $\mathcal{L}^{wav}$ between the prediction and target waveform to match the overall shape and the phase. Inspired by recent studies in speech synthesis and enhancement~{\cite{yamamoto2020parallel,kong2020hifi,defossez2020real}}, a multi-resolution STFT loss (i.e., $\mathcal{L}^{stft}$) is also introduced to improve the perceptually-related auditory quality, which is a combination of the spectral convergence loss and the logarithmic magnitude spectral L1 loss with different FFT analysis parameters. In our experiments, the number of FFT bins is set to $\{512,1024,2048\}$, the hop length is set to ${\in}\{50,120,240\}$, and the window size is set to $\{240,600,1200\}$. 
     
     
    For adversarial training, we introduce the multi-scale STFT-based (MS-STFT) adversarial training to capture different patterns in speech signals~{\cite{kumar2019melgan,defossez2022high}}. Following previous studies, the relativistic average least-square loss (RaLSGAN) proposed in ~{\cite{jolicoeur2018relativistic}} is adopted to stabilize the competitive relationship between multiple discriminators $D_i$ and the generator $G$, which can be formulated as:
    \begin{gather}  
    \mathcal{L}^{adv}_{D}= \mathbb{E}_{\bf s}\left [(D(s)-1)^{2}  \right ] +\mathbb{E}_{\widetilde{\bf s}}\left [(D(\widetilde{\bf s})+1)^{2}  \right ], \\
        \mathcal{L}^{adv}_{G}= \mathbb{E}_{\widetilde{\bf s}}\left [(D(\widetilde{\bf s})-1)^{2}  \right ] +\mathbb{E}_{{\bf s}}\left [(D({\bf s})+1)^{2}  \right ]
    \end{gather}
    where the generated HR time-domain signal is denoted by ${\bf \widetilde s}$ and the target is denoted by ${\bf s}$. Additionally, the feature match loss~{\cite{kumar2019melgan}} $\mathcal{L}^{feat}_{G}$ is also adopted to minimize the L1 distance between the feature maps of the discriminator’s internal outputs for real HR speech and those for the corresponding generated speech
    
    The overall loss of the generator is a weighted sum of the waveform loss, the multi-resolution STFT loss and the adversarial-related loss, which can be finally given by:
    \begin{gather}  
        \mathcal{L}_{G}=\lambda_{\rm wav} \mathcal{L}^{wav} + \lambda_{\rm sfft}\mathcal{L}^{stft} + \mathcal{L}_{adv}+ \lambda_{\rm feat} \mathcal{L}_{feat} ,
    \end{gather}
     where $\lambda_{\rm wav}$, $\lambda_{\rm stft}$ and $\lambda_{\rm feat}$ are set to 100, 0.5 and 10, respectively.

        \renewcommand\arraystretch{1.05}
        \begin{table*}[t!]
            \setcounter{table}{1}
            \caption{Objective measures with other SOTA baselines }
            \centering
            \scalebox{0.85}{
                \begin{tabular}{l|cc|cccc|ccc}
                    \hline
                    \multirow{2}*{Models} &{Para.} &{MACs} &\multicolumn{4}{c|}{\textbf{8-48kHz}}   &\multicolumn{3}{c}{\textbf{16-48kHz}}\\
                    \cline{4-7} \cline{8-10}
                    &(M) &(G/s) &{SegSNR(dB)$\uparrow$}  &{LSD$\downarrow$}    &PESQ$\uparrow$ &POLQA$\uparrow$ &{SegSNR(dB)$\uparrow$}   &{LSD$\downarrow$}  &POLQA$\uparrow$ \\ 
                    \hline
                Cubic Spine &\makecell[c]{--} & \makecell[c]{--}  &17.98         &2.22       &3.85      &3.87 &23.79 &1.13 &4.36  \\ 
                    \hline
                    AudioUnet~{\cite{kuleshov2017audio}} &35.78 & 85.01 &19.35 &1.20 &3.56 &3.68 &24.48 & 0.97 &3.96 \\ 
                TF-Net~{\cite{lim2018time}} & 22.60 & 132.09  &22.45 &0.96 &3.76 &3.92 &26.89 &0.72 &4.18  \\ 
                    SEANet~{\cite{li2021real}} & 4.97 & 15.63  &25.94 &0.79 &3.99 &4.25 &30.96 &0.54 &4.56  \\ 
                    AERO~{\cite{mandel2023aero}} &20.45 & 25.87  &24.32 &0.82 &3.92 &4.26 & 28.98&0.59 &4.54  \\ 
                    \hline
                    \multicolumn{8}{c}{\textbf{Our proposed method}} \\     
                    \hline
                    BAE-Net-lite &\textbf{0.57} & \textbf{0.057}  &25.77 &0.80 &3.94 &4.19 &31.44 &\textbf{0.49} &4.58  \\ 
                BAE-Net &3.08 & 0.31  &\textbf{26.15} &\textbf{0.78} &\textbf{3.96} &\textbf{4.30} &\textbf{32.13} &0.50 &\textbf{4.62}  \\ 
                    ~- BGM &3.07 & 0.31  &21.23 &0.86 &3.74 &3.95 &29.06 &0.59 &4.39  \\ 
                    ~~~~- Inter. &2.99 & 0.30  &20.94 &0.80 &3.67 &3.89 &28.41 &0.63 &4.37  \\ 
                \hline

                \end{tabular}
            }
            \label{tbl:results}
        \end{table*}
     \vspace{-0.3cm}
    \section{Experiments\label{Section3}}
    \label{Sec3}
 \vspace{-0.2cm}
    \subsection{Dataset and Implementation\label{Section31}}
    \vspace{-2mm} 
    Due to the limited network complexity, BAE-Net focuses on the speech super-resolution task instead of music super-resolution. We evaluate our model on speech signals taken from the widely used VCTK dataset~{\cite{veaux2017cstr}}, which contains around 44 hours of clean speech with the sampling rate 48 kHz from 110 speakers. The training set includes 108 speakers, and the rest 2 unseen speakers are chosen for testing, which is the same as the testset in~{\cite{valentini2016investigating}}. 
 
    When comparing the performance of our model with other baselines which only support fixed up-sampling ratios, we set two up-sampling settings: 8-48 kHz and 16-48 kHz sampling rate. For adaptive-bandwidth training, the effective bandwidth of input signal is variable, where the high frequency cutoffs are uniformly sampled in the ranges of 8kHz to 48kHz sampling rate. The input LR speech is produced from the ground truth speech by means of low-pass filtering which can be performed on-the-fly during training, while the sampling rate remains 48 kHz. This is because that in practical RTC scenarios, the sampling rate of the pipeline is typically fixed while the effective bandwidth of captured speech frequently fluctuates.
    The Hanning window with the length 32 ms is selected, with 50\% overlap between consecutive frames. The 1536-point STFT is utilized resulting 769-dimension spectral features.
    All the models are optimized using Adam~{\cite{kingma2014adam}} with the learning rates of 2e-5 and 1e-5 for the generator and discriminators, repectively. \textbf{The processed samples are available online.}{\footnote{https://github.com/yuguochencuc/BAE-Net}
\vspace{-0.3cm}   
    \subsection{Baselines\label{Section32}}
    \vspace{-0.2cm}   
    We implement several state-of-the-art waveform-based and spectrum-based approaches as the benchmarking references for evaluation of fixed-upsampling ratio scenarios, including: a) simple bicubic interpolation (\textbf{Cubic Spline}); b) a time-domain based convolutional U-net (\textbf{AudioUnet})~{\cite{kuleshov2017audio}}; c) a time-frequency domain fusion network (\textbf{TF-Net})~{\cite{lim2018time}}; d) a wave-to-wave fully convolutional model (\textbf{SEANet})~{\cite{li2021real}}; e) a complex-spectrum based audio super-resolution model (\textbf{AERO})~{\cite{mandel2023aero}}. We also conduct the ablation study to investigate the effect of the proposed Band-Guided Masking (BGM) module and the information interaction pathway (Inter.).
    
\vspace{-0.3cm}   
    \subsection{ Evaluation Metrics\label{Section32}}
    \vspace{-1mm}  
        
         For 8-48 kHz bandwidth extension, four objective signal-based evaluation indicators are used to measure the quality of the reconstructed speech, including segmental signal-to-noise ratio (\textbf{SegSNR}), log-spectral distance (\textbf{LSD}), perceptual evaluation of speech quality (\textbf{PESQ})~{\cite{hu2007evaluation}}, and perceptual objective listening quality assessment (\textbf{POLQA})~{\cite{beerends2013perceptual}}. Note that we downsample the generated 48 kHz full-band speech with the 16kHz sampling rate to compute PESQ, because only wide-band and narow-band speech can be evaluated using the PESQ metric. For 16-48 kHz bandwidth extension, we employ SegSNR, LSD and POLQA to evaluate the speech quality. Note that the network weights of BAE-Net is unaltered in all experiments.

    \vspace{-0.2cm}      
        \section{Results and Analysis\label{Section4}}
        \label{Sec4}
\vspace{-0.2cm}

\begin{table}[!t]
\small
\centering
\vspace{-0.4cm}
\caption{Evaluation on the fluctuated effective bandwidth}
\resizebox{0.7\columnwidth}{!}{%
\begin{tabular}{l|c|cc} 
    \toprule
    Setting & method & LSD  $\downarrow$ & POLQA $\uparrow$ \\
    \midrule
    \multirow{3}*{10-48 kHz} &unprocessed & 2.55  &3.96 \\
    \cline{2-4} 
    & AERO~{\cite{mandel2023aero}} & 0.82 & 4.26\\
    \cline{2-4} 
    & BAE-Net & \textbf{0.69} & \textbf{4.38}\\
    \cline{1-4} 
        \multirow{3}*{14-48 kHz} &unprocessed & 2.03 &4.40\\
            \cline{2-4} 
    & AERO~{\cite{mandel2023aero}} & 0.69 & 4.42\\
    \cline{2-4} 
    & BAE-Net  & \textbf{0.56} & \textbf{4.54}\\
    \cline{1-4} 
        \multirow{3}*{20-48 kHz} &unprocessed & 1.05 &4.58\\
            \cline{2-4} 
    & AERO~{\cite{mandel2023aero}} & 0.51 & 4.57\\
    \cline{2-4} 
    & BAE-Net & \textbf{0.42}  & \textbf{4.64}\\
    \cline{1-4} 
        \multirow{3}*{24-48 kHz} &unprocessed & 0.08 &4.61\\
            \cline{2-4} 
    & AERO~{\cite{mandel2023aero}} & \textbf{0.03} & 4.65\\
    \cline{2-4} 
    & BAE-Net & 0.04  & \textbf{4.68}\\
    \bottomrule
    \end{tabular}
}
\label{tbl:variable_band_results}
\vspace{-0.4cm}
\end{table}

        \subsection{Comparison on the Fixed Upsampling Ratios\label{Section42}}
        \vspace{-2mm}  
        We first compare the objective performance of the proposed methods with other state-of-the-art (SOTA) baselines for two fixed upsampling ratio scenarios including 8 kHz to 48 kHz and 16 kHz to 48 kHz. As presented in Table~{\ref{tbl:results}}, we also provide detailed model complexity comparisons in terms of the number of parameters (Para.(M)), and the multiply-accumulate operations (MACs). From Table~{\ref{tbl:results}}, we have the following observations. 
        First, for the 8-48 kHz and 16-48 kHz tasks, compared with previous waveform-based and spectrum-based baselines, BAE-Net achieves consistently better performance in term of most metrics.
         For the 8-48 kHz and 16-48 kHz tasks, BAE-Net outperforms AERO by average 2.49 dB in SeSNR and 0.07 in POLQA, while having about 80 times fewer MACs. 
         Second, although incurring much fewer network parameters and MACs, BAE-Net-lite achieves moderate scores and is competitive with existing methods as well as BAE-Net. This verifies that BAE-Net-lite provides an applicable BWE technique for edge devices.
        Third, we investigate the impact of the BGM module and the interaction pathway. It can be observed that the performance of BAE-Net significantly degrades when BGM and Inter. are not incorporated, particularly without the BGM module. 
  

 \begin{figure}[t]
    \centering
    \centerline{\includegraphics[width=0.9\columnwidth]{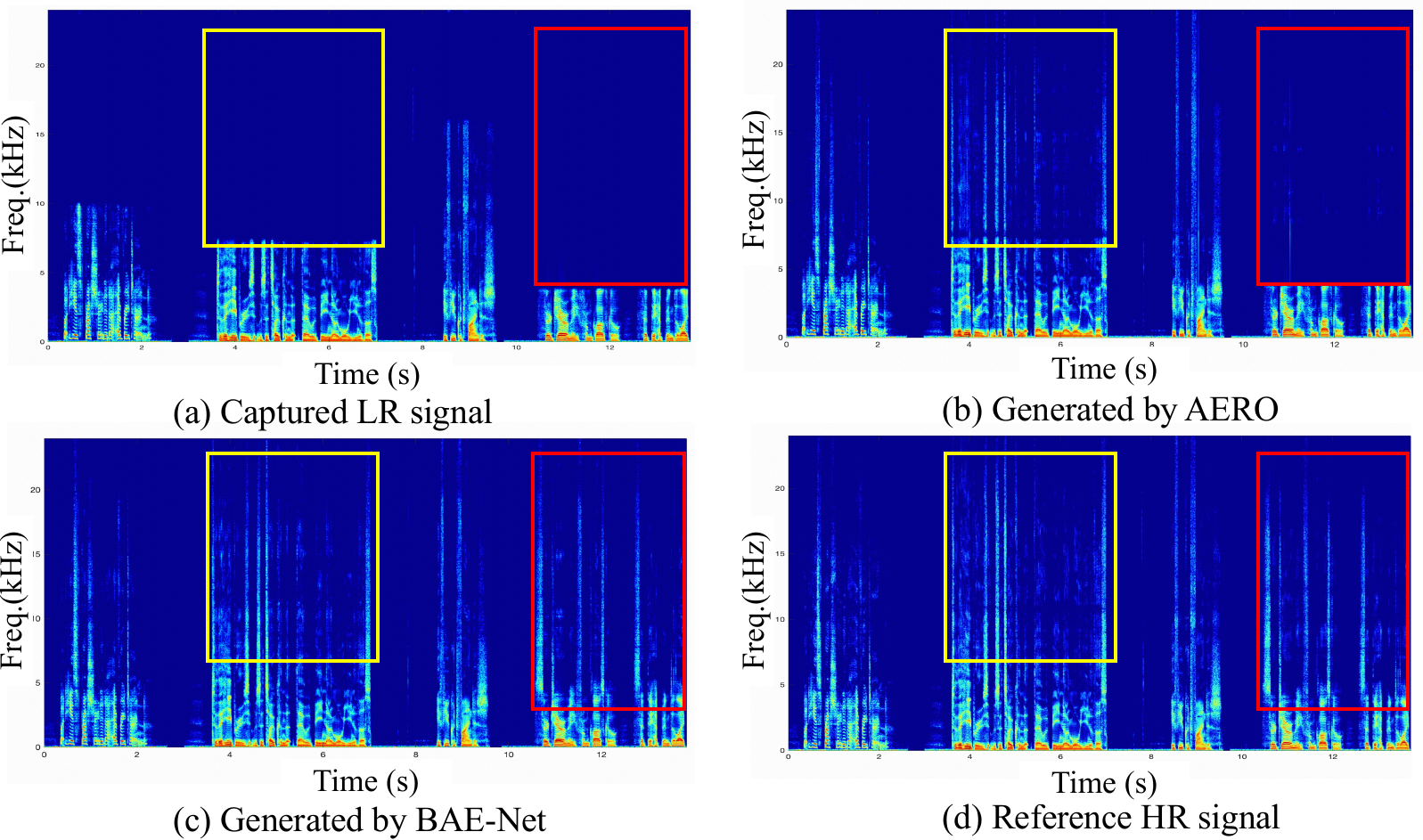}}
    \vspace{-0.3cm}
    \caption{Visualization of spectrograms of captured LR speech, generated HR speech by AERO and BAE-Net, and the HR reference.}
    \label{fig:result_example}
    \vspace{-0.5cm}
    \end{figure}
    
        \vspace{-4mm}
        \subsection{Evaluation on Variable Upsampling Ratios \label{Section42}}
        To verify the robustness of the proposed method in real unknown-bandwidth scenarios, we conduct evaluations on variable effective bandwidths. In order to accommodate the fixed upsampling rates (i.e., 8-48 kHz, 16-48 kHz and 24-48 kHz) supported by our re-implemented AERO, the test speech signals are resampled to the nearest sampling rate that approximates the original effective bandwidth, i.e., resampling the input signals from 10 kHz to 8kHz, from 14 kHz to 16kHz, and from 20 kHz to 24kHz.

        As illustrated in Table~{\ref{tbl:variable_band_results}}, BAE-Net achieves competitive performance improvement compared with AERO and the unprocessed LR signals at each sampling rate. 
        To provide visual evidence, we present an example of spectrograms of LR signal with fluctuated effective bandwidth and generated HR signals in Figure~{\ref{fig:result_example}}. Due to the fixed up-sampling ratio supported by AERO, the input LR signal for AERO is resampled at fixed sampling rate 16 kHz. It can be observed that although the effective bandwidth frequently fluctuates in the utterance, BAE-Net demonstrates the efficacy and superiority in handing the unknown bandwidth scenarios. In contrast, AERO fails to reconstruct the high-frequency components within the red box, suffering from significant performance degradation.


        %
        %
  \vspace{-0.3cm}
        \section{Conclusions\label{Section5}}
  \vspace{-0.3cm}
        \label{Sec5}

  In this paper, we introduce BAE-Net, a dual-stream low-complexity network designed to address adaptive speech bandwidth extension in practical scenarios where the effective bandwidth of captured audio fluctuates frequently. To be specific, a magnitude inpainting stream and a phase refinement network are devised to collaboratively facilitate the recovery of magnitude and phase information in the missing high-frequency speech components.
    Additionally, a lightweight and streaming network BAE-Net-lite is also introduced for edge-device applications and achieves comparable performance. Experimental results on the fixed up-sampling ratios and speech with fluctuated effective bandwidth demonstrate that the proposed method achieves state-of-the-art performance over previous competitive systems with a relatively less computation cost and a smaller model size.
        
\vspace{-3mm}

\bibliographystyle{IEEEbib}
\begin{spacing}{0.9} 
    \bibliography{myrefs} 
\end{spacing} 

\end{document}